%% file: paper.tex
\documentclass[USenglish,thm-restate]{lipics-v2021}

\hideLIPIcs  

\nolinenumbers 

\input{packages}
\input{commands}

\input{environments}

\input{tikz_commands}

\newcounter{cor-memory-safety}
\newcounter{thm-soundness-graph-construction-main-text}
\newcounter{thm-termination}

\newcommand{\report}[1]{}
\newcommand{\arxiv}[1]{}

\title{Automated Termination Proofs for \tool{C} Programs with Lists\\(Short WST Version)}
\titlerunning{Automated Termination Proofs for \tool{C} Programs with Lists}
\author{Jera Hensel}{\and
  \url{https://verify.rwth-aachen.de/jhensel/}}{hensel@informatik.rwth-aachen.de}{https://orcid.org/0000-0003-2852-9830}{}
\author{Jürgen Giesl}{LuFG Informatik 2, RWTH Aachen University, Aachen, Germany \and \url{https://verify.rwth-aachen.de/giesl/}}{giesl@informatik.rwth-aachen.de}{https://orcid.org/0000-0003-0283-8520}{}
\authorrunning{J.\ Hensel, J.\ Giesl}

\Copyright{Jera Hensel and Jürgen Giesl}

\relatedversiondetails{\vspace*{-.2cm}See \cite{report}. Full version, including all
  proofs}{https://arxiv.org/abs/2305.12159} 

\funding{funded by the Deutsche Forschungsgemeinschaft (DFG, German Research Foundation) -
  235950644 (Project GI 274/6-2)}


\keywords{\vspace*{-.2cm}Termination Analysis, \tool{C} Programs, Lists, Symbolic Execution}

\begin{document}

\maketitle

\begin{abstract}\input{abstract}\end{abstract}
\input{introduction}
\input{domain}

\input{inferring_inv}

\input{evaluation}



\input{paper.bbl}

\end{document}

%% file: packages.tex
\usepackage{amsmath,amssymb} 
\let\originallhook\lhook
\usepackage{mnsymbol}
\let\lhook\originallhook

\usepackage{varioref}
\usepackage{xcolor} 
\usepackage{listings} 
\usepackage{lipsum}	
\usepackage[pangram]{blindtext}
\usepackage[capitalize]{cleveref}
\usepackage{mathtools}
\usepackage[normalem]{ulem}
\usepackage{placeins}
\usepackage{cprotect}
\usepackage{wrapfig}
\usepackage{subcaption}
\usepackage{fancyvrb}
\usepackage{boxedminipage}
\usepackage{array}
\usepackage{stmaryrd}
\usepackage[shortlabels]{enumitem}
\usepackage{fancyvrb}
	
\usepackage{graphicx}
\PassOptionsToPackage{usenames,dvipsnames,svgnames,table}{xcolor}
\usepackage{ wasysym }
\usepackage{tikz}
\usepackage{pgfplots}
\usetikzlibrary{arrows}
\usetikzlibrary{patterns}
\usetikzlibrary{shapes,arrows,calc,arrows}
\usetikzlibrary{automata,arrows,decorations.pathmorphing,positioning,fit}
\usetikzlibrary{trees,shapes}
\usetikzlibrary{decorations.pathreplacing}
\usetikzlibrary{calc}
\usetikzlibrary{intersections}
\usetikzlibrary{backgrounds}
\usetikzlibrary{fadings}
\usetikzlibrary{tikzmark}

\usepackage{calc}
\usepackage{tkz-euclide}
\usepackage{catchfilebetweentags}
\usepackage{afterpage}
\usepackage{url}
\usepackage{todonotes}
\usepackage{xstring}
\usepackage{nameref}
\usepackage{alphalph}
\usepackage{relsize}

\usepackage[utf8]{inputenc} 
\usepackage{algorithm2e} 
\usepackage{scrextend}
\usepackage{refcount} 

\usepackage{forloop}
\usepackage{marvosym}

%% file: commands.tex
\renewcommand{\emptyset}{\varnothing}

\newcommand{\code}[1]{\texttt{#1}}
\newcommand{\tool}[1]{\textsf{#1}}
\newcommand{\sfC}{\tool{C}}
\newcommand{\LLVM}{\tool{LLVM}\xspace}

\newcommand{\aprove}{\tool{AProVE}}

\newcommand{\PP}{\mathcal{P}}

\newcommand{\Ids}{\mathcal{V}_{\PP}}

\newcommand{\N}{\mathbb{N}}

\newcommand{\Prog}{\mathcal{P}}

\newcommand{\ERROR}{\mathit{ERR}}

\newcommand{\oldcom}[1]{}

\newcommand{\codevar}[1]{\ensuremath{\mathtt{#1}}}

\newcommand{\pointsto}[1][]{\hookrightarrow_{#1}} 
\newcommand{\pointstorec}[2][]{\xhookrightarrow{#2}_{#1}} 

\newcommand{\domain}{\mathop{domain}} 

\newcommand{\LI}{\mathit{LI}}
\newcommand{\PT}{\mathit{PT}}

\newcommand{\KB}{\mathit{KB}}

\newcommand{\LV}{\mathit{LV}}

\newcommand{\alloc}[2]{\ensuremath{\llbracket{}#1,\,#2\rrbracket}}

\newcommand{\AL}{\mathit{AL}}

\newcommand{\Vsym}{\mathcal{V}_{\mathit{sym}}}

\newcommand{\partialfunctionmap}[0]{\rightharpoonup}

\newcommand{\cmpFor}{\code{cmpF}}
\newcommand{\bodyFor}{\code{bodyF}}

\newcommand{\initPtr}{\code{initPtr}}

\newcommand{\GraphInitEntryZero}{A}
\newcommand{\GraphInitCmpZero}{B}
\newcommand{\GraphInitCmpOne}{C}
\newcommand{\GraphInitCmpOneRefTrue}{D}
\newcommand{\GraphInitCmpOneRefFalse}{E}
\newcommand{\GraphInitCmpTwoTrue}{F}
\newcommand{\GraphInitBodyZero}{G}
\newcommand{\GraphInitBodyOne}{H}
\newcommand{\GraphInitBodySix}{J}
\newcommand{\GraphInitBodyTen}{K}
\newcommand{\GraphInitCmpZeroA}{L}
\newcommand{\GraphInitCmpZeroB}{M}
\newcommand{\GraphInitCmpZeroGen}{O}

\newcommand{\StateA}{s}
\newcommand{\StateB}{{s'}}
\newcommand{\StateGen}{{\overline{s}}}

\newcommand{\seqs}

\renewcommand{\arraystretch}{1.2}

%% file: environments.tex
\DeclareMathAlphabet
{\mathttit}{OT1}{lmtt}{l}{it}

\newlist{caselist}{enumerate}{10}
\setlist[caselist]{wide,labelindent=0pt,itemsep=1ex,topsep=1ex,parsep=0pt,leftmargin=0pt,%
 labelwidth=*,labelsep=1ex,align=caselabel,label*=.\arabic*,}
\setlist[caselist,1]{label=\arabic*}
\newcommand\thecaselabeltext{}
\newcommand{\thecaselabelword}{Case}
\SetLabelAlign{caselabel}{\uline{\thecaselabelword~#1\thecaselabeltext}.\hfil}
\crefname{caselisti}{Case}{Cases}
\crefname{theorem}{Thm.}{Thms.}
\Crefname{theorem}{Theorem}{Theorems}
\crefname{lemma}{Lem.}{Lems.\ }
\Crefname{lemma}{Lemma}{Lemmas}
\crefname{subformula}{subformula}{subformulas}
\Crefname{subformula}{Subformula}{Subformulas}
\crefname{corollary}{Cor.}{Corollaries}
\Crefname{corollary}{Corollary}{Corollaries}
\crefname{definition}{Def.}{Def.}
\Crefname{definition}{Definition}{Definitions}
\crefname{proofcondition}{Cond.}{Conditions}
\Crefname{proofcondition}{Condition}{Definitions}
\crefname{appendix}{App.}{App.}
\Crefname{appendix}{Appendix}{Appendixes}
\crefname{page}{page}{pages}
\Crefname{page}{Page}{Pages}
\creflabelformat{subformula}{#2(#1)#3}
\crefname{condition}{Cond.}{Conditions}
\Crefname{condition}{Condition}{Definitions}
\crefname{section}{Sect.}{FIXME}
\Crefname{section}{Section}{FIXME}
\crefname{eqstatement}{}{}
\Crefname{eqstatement}{}{}
\newcounter{i}
\forLoop{1}{10}{i}{\crefalias{caselist\roman{i}}{caselisti}}

\newlist{labeledenumerate}{enumerate}{10}
\setlist[labeledenumerate]{leftmargin=*,labelindent=1em,topsep=5pt}

%% file: tikz_commands.tex
\tikzstyle{eval-edge}=[-stealth]
\tikzstyle{gen-edge}=[-stealth]
\tikzstyle{int-gen-edge}=[-stealth]
\tikzstyle{ca-edge}=[-stealth]
\tikzstyle{fs-edge}=[-stealth]
\tikzstyle{omit-edge}=[-stealth, dotted]

\usetikzlibrary{shapes,positioning,calc}
\usepackage{wrapfig}
\usetikzlibrary{shapes,arrows,calc}
\tikzstyle{ptr}  = [draw, -latex']
\tikzstyle{ptrvar} = [rectangle, draw, text height=3mm, text width=3mm, text centered, node distance=3cm, inner sep=0pt]
\tikzstyle{data} = [rectangle split, rectangle split horizontal, rectangle split parts=2, draw, text centered, text width=0.4cm]
\tikzstyle{mem16} = [rectangle split, rectangle split horizontal, rectangle split parts=16, draw, text centered, text width=0.4cm]

\tikzset{
    invisible/.style={opacity=0},
    visible on/.style={alt={#1{}{invisible}}},
    alt/.code args={<#1>#2#3}{%
      \alt<#1>{\pgfkeysalso{#2}}{\pgfkeysalso{#3}} 
    },
  }
\tikzset{onslide/.code args={<#1>#2}{%
  \only<#1>{\pgfkeysalso{#2}} 
}}

\pgfdeclaremetadecoration{middlezigzag}{straight}{
    \state{straight}[switch if less than=\pgfmetadecorationsegmentlength to final,
                  width=\pgfmetadecoratedpathlength/2 - \pgfmetadecorationsegmentlength/2 ,
                  next state=zigzag] {
        \decoration{curveto}
    }
    \state{zigzag}[width=\pgfmetadecorationsegmentlength,
                   next state=final] {
        \decoration{zigzag}
    }
    \state{final}{
        \decoration{curveto}
        \beforedecoration{\pgfpathmoveto{\pgfpointmetadecoratedpathfirst}}
    }
}

\tikzset{
    middle zigzag/.style={
        decorate,
        decoration={
            middlezigzag,
            meta-segment length=#1,
            segment length=0.5cm,
			amplitude=1.5pt
        }
    },
    middle zigzag/.default=1cm
}

%% file: abstract.tex
There are many techniques and tools for termination
of \sfC{} programs, but up to now 
they were not very powerful
for 
termination proofs of 
programs whose
termination depends on recursive data structures like lists.
We present the
first approach that extends powerful techniques for termination analysis of \sfC{} programs (with
memory allocation and explicit pointer arithmetic) to lists.

%% file: introduction.tex
\section{Introduction}
\label{sect:introduction}

In \cite{ASV19,JLAMP18,LLVM-JAR17}, we intro\-duced an approach for
termination analysis of
\sfC{} that also handles programs whose termination relies on the relation
between
allocated memory and the data stored at such addresses.
This approach is implemented in our tool \aprove{} \cite{JAR17-AProVE}.
Instead of analyzing \sfC{} directly, \aprove{} compiles the program to \LLVM{}.
Then it constructs a (finite) symbolic execution graph (SEG) such that every program run
corresponds to a path through
the SEG.
\aprove{} proves memory safety during the construction of the SEG to ensure
absence of undefined behavior (which would also
allow non-termination). Afterwards,
the SEG is transformed into an integer transition system (ITS) such
that all paths through the SEG (and hence, the \sfC{} program)
are terminating if the ITS is terminating. To analyze termination of 
 ITSs, 
\aprove{} applies standard techniques and tools.
However, like other termination tools  for \sfC{},
\noindent
  up to now
   \aprove{} supported dynamic data structures only in a very restricted way.

\begin{wrapfigure}[13]{r}{5.6cm}
\vspace*{-.3cm}
\scriptsize
\begin{minipage}{5.2cm}
  \begin{Verbatim}[commandchars=\\\{\}]
struct list \{
  unsigned int value;
  struct list* next;   \};

int main() \{
  \textcolor{black!55}{// initialize length}
  unsigned int n = nondet_uint();
  \textcolor{black!55}{// initialize list of length n}
  struct list* tail = NULL;
  struct list* curr;
  for (unsigned int k = 0; k < n; k++) \{
    curr = malloc(sizeof(struct list));
    curr->value = nondet_uint();
    curr->next = tail;
    tail = curr;                       \}
  \textcolor{black!55}{// traverse list}
  struct list* ptr = tail;
  while(ptr != NULL) \{
    ptr = *((struct list**)((void*)ptr +
          offsetof(struct list, next)));\}\}
  \end{Verbatim}
\end{minipage}
\end{wrapfigure}
In the program on the side, \code{nondet\_uint}  returns a random unsigned integer. The
\code{for} loop creates a list of \code{n} random numbers if $\code{n}> 0$ and the\linebreak
\code{while} loop traverses this list
via poin\-ter arithmetic.
We introduce a novel technique which can
analyze termination of
such \sfC{} programs on lists, i.e., it can both handle
the
 access by \code{curr->next} (when initializing the list) and by pointer
  arithmetic (when traversing the list).

  Our technique infers \emph{list invariants} via symbolic execution. These invariants
express
all properties that are crucial for
memory safety and termination.
In our example, the list invariant contains the 
information that dereferencing the \code{next} pointer
in the \code{while} loop
is  safe and that one finally reaches
the null pointer.

To ease the presentation, in this paper we treat integer types as unbounded.
Moreover, we assume that a program consists of a
single non-recursive function
and that values may be stored at any address.
Our approach can also deal with bitvectors, data
alignments, and programs with arbitrary many (possibly recursive) functions, see
\cite{ASV19,JLAMP18,LLVM-JAR17} for details.
However,  so far only lists without sharing can be handled by our new technique. Extending it to more
general recursive data structures is one of the main challenges for future
work.

%% file: domain.tex
\section{Abstract States for Symbolic Execution}\label{sect:domain}

\begin{wrapfigure}[18]{r}{6cm}
\vspace*{-.3cm}
\begin{boxedminipage}{6cm}
\scriptsize
\begin{Verbatim}[commandchars=\\\{\}]
list = type \{ i32, list* \}

define i32 @main() \{ ...
\cmpFor:
  \tiny{\textcolor{black!55}{k < n}}
  0: k = load i32, i32* k_ad
  1: kltn = icmp ult i32 k, n
  2: br i1 kltn, label \bodyFor, label \initPtr
\bodyFor:
  \tiny{\textcolor{black!55}{curr = malloc(sizeof(struct list));}}
  0: mem = call i8* @malloc(i64 16)
  1: curr = bitcast i8* mem to list*
  \tiny{\textcolor{black!55}{curr->value = nondet_uint();}}
  2: nondet = call i32 @nondet_uint()
  3: curr_val = getelementptr list,
                list* curr, i32 0, i32 0
  4: store i32 nondet, i32* curr_val
  \tiny{\textcolor{black!55}{curr->next = tail;}}
  5: tail = load list*, list** tail_ptr
  6: curr_next = getelementptr list,
                 list* curr, i32 0, i32 1
  7: store list* tail, list** curr_next
  \tiny{\textcolor{black!55}{tail = curr;}}
  8: store list* curr, list** tail_ptr
  \tiny{\textcolor{black!55}{k++}}
  9: kinc = add i32 k, 1
  10:store i32 kinc, i32* k_ad
  11:br label \cmpFor     ...                 \}
\end{Verbatim}
\end{boxedminipage}
\vspace{-.05cm}
\end{wrapfigure}

\noindent
The \LLVM{} code 
for the \code{for} loop is given on the side.
To ease readability, we omitted instructions and keywords that are irrelevant for our
presentation, renamed variables, and wrote \code{list} instead of \code{struct.list}.
The code consists of several \emph{basic blocks} including \cmpFor{}
 and \bodyFor{} 
(for the  
comparison and the body of the \code{for} loop).

We now recapitulate  the \emph{abstract states}
of
\cite{LLVM-JAR17}  used for symbolic
execu\-tion and extend  them by a component $\LI$ for list
invariants.
The first component is a \emph{program position}
(\codevar{b}, $i$), indicating that instruction $i$ of block \codevar{b} is executed
next. 

The second component is a partial injective
function $\LV\colon\Ids\partialfunctionmap\Vsym$, which maps \emph{\underline{l}ocal program \underline{v}ariables} $\Ids$ of the program
$\PP$ to
an infinite set  $\Vsym$ of symbolic variables with $\Vsym
\cap \Ids = \emptyset$.

The third component is a set $\AL$ of allocations  $\alloc{v_1}{v_2}$ with
$v_1, v_2 \in \Vsym$,  indicating that $v_1 \le v_2$
and all addresses between $v_1$ and $v_2$ are allocated. 

The fourth and fifth components  $\PT$ and $\LI$ model the memory. $\PT$ contains
``\underline{p}oints-\underline{t}o'' entries $v_1 \pointsto[\codevar{ty}] v_2$ where $v_1,v_2 \in \Vsym$ and
$\codevar{ty}$ is an \LLVM{} type, meaning that the address $v_1$ of type $\codevar{ty}$
\underline{p}oints \underline{t}o $v_2$. 
The set $\LI$ of \emph{\underline{l}ist \underline{i}nvariants} (which is new
compared to \cite{LLVM-JAR17})  contains invariants
$v_{\mathit{ad}} \pointstorec[\codevar{ty}]{v_\ell} [(\mathit{off}_i: \codevar{ty}_i:
  {v_i..\hat{v}_i})]_{i=1}^m$ where $m\!\in\!\N_{>0}$,
  $v_{\mathit{ad}},v_\ell,v_i,\hat{v}_i \in
\Vsym$, $\mathit{off}_i \in \N$
for all $1 \leq i \leq m$, $\codevar{ty}$ and
$\codevar{ty}_i$ are \LLVM{} types for all $1 \leq i \leq m$, and there is exactly
one ``recursive field''
$1 \leq j \leq m$ such that $\code{ty}_j = \code{ty*}$.
Such an invariant represents a
\code{struct} \code{ty} with $m$ fields that corresponds to a recursively defined list of
length $v_\ell$.
Here, $v_{\mathit{ad}}$ points to the first list element, the $i$-th field starts at address $v_{\mathit{ad}} + \mathit{off}_i$ 
and has type $\codevar{ty}_i$,
and the values of the $i$-th fields of
the first and last list element
are $v_i$ and $\hat{v}_i$, respectively.
For example, the list invariant \eqref{example list invariant} represents all lists of length $x_{\ell}$
and type \code{list} whose elements store a 32-bit integer in their first field and the pointer to the next element in their second field
with offset $8$. The first list element starts at address $x_\code{mem}$,
the second  starts at ad\-dress $x_\code{next}$, and the last element
contains the null pointer.
Moreover, the first  element stores the integer value $x_\code{nd}$ and the last list
element stores  the  integer $\hat{x}_\code{nd}$. 
\begin{equation}
  \label{example list invariant}
x_{\code{mem}} \pointstorec[\code{list}]{x_{\ell}} [(0: \code{i32}:
  {x_\code{nd}}..\hat{x}_{\code{nd}}), (8: \code{list*}: x_{\code{next}}..0)]
\end{equation}
\noindent
The last component  is a \emph{\underline{k}nowledge \underline{b}ase} $\KB$
of formulas that express
arithmetic properties of $\Vsym$. 
A special state $\ERROR$ is reached if we cannot prove absence of undefined
behavior.

As an example, the abstract state \eqref{example state}
represents concrete states at the beginning of the
block \cmpFor{}, where the program variable \code{curr} is assigned the symbolic variable
$x_{\code{mem}}$, the allocation $\alloc{x_{\code{k\_ad}}}{x_{\code{k\_ad}}^\mathit{end}}$
consisting of $4$ bytes stores the value $x_{\code{kinc}}$,
and $x_{\code{mem}}$ points to
the first element of a list of length $x_\ell$ (equal to $x_{\code{kinc}}$)
that satisfies the list invariant \eqref{example list invariant}.
 (This state will later
 be obtained during the symbolic execution, see State $O$ in \cref{fig:ForLoopMerging}.)
\begin{equation}
  \label{example state}
 \parbox{6cm}{
 \begin{tikzpicture}[node distance = \ydist and \xdist]
\scriptsize
\tikzstyle{state}=[inner sep=2pt, font=\scriptsize, draw]
\node[state, align=left] (1) 
{$(\cmpFor, 0),\;
  \{ \code{curr} = x_{\code{mem}}, \,
     \code{kinc} = x_{\code{kinc}}, \,
     ...\}, \;
  \{ \alloc{x_{\code{k\_ad}}}{x_{\code{k\_ad}}^\mathit{end}}, \,
     ...\}, \;
  \{ x_{\code{k\_ad}} \pointsto[\code{i32}] x_{\code{kinc}}, \,
     ...\}, \;$\\
 $\{ x_{\code{mem}} \pointstorec[\code{list}]{x_{\ell}} [(0: \code{i32}: x_{\code{nd}}..\hat{x}_{\code{nd}}), (8: \code{list*}: x_{\code{next}}..0)]\}, \;
  \{ x_{\code{k\_ad}}^\mathit{end} = x_{\code{k\_ad}} + 3, \,
     x_{\ell} = x_{\code{kinc}}, \,
     ...\}
 $
};
\end{tikzpicture}}
\end{equation}

%% file: inferring_inv.tex
\section{Symbolic Execution with List Invariants}
\label{sect:symbexec}

\input{graph_init_first}
Symbolic execution starts with a state $\GraphInitEntryZero$
at the first instruction of the first block
(\code{entry} in
our\linebreak example), see
Fig.\ \ref{fig:ForLoopFirstIteration}.
Dotted arrows indicate that we omitted some  steps. For every
state, we\linebreak perform symbolic execution by applying  
the corresponding inference rule 
to compute
its successor(s) and repeat this until all paths end in return states.
Such an SEG is  \emph{complete}.

In our example,
the \code{entry} block comprises the first three lines of the \sfC{} program and
the\linebreak initialization of the pointer to the loop variable $\code{k}$:
First,  a non-deterministic integer is assigned to \code{n}, i.e.,
 $(\code{n} = v_\code{n}) \in \LV^{\GraphInitCmpZero}$, where
 $v_\code{n}$ is not restricted.
  Moreover,  memory for the
pointers \code{tail\_ptr} and \code{k\_ad} is allocated and they
point to \code{tail = NULL} and \code{k = 0}, respectively
($\code{tail\_ptr} = v_{\code{tp}}$
and $\code{k\_ad} = v_{\code{k\_ad}}$ 
with $(v_{\code{tp}} \pointsto[\code{list*}] 0),
(v_{\code{k\_ad}} \pointsto[\code{i32}] 0)
\in \PT^\GraphInitCmpZero$). For
simplicity, in \cref{fig:ForLoopFirstIteration}
we use concrete values directly instead of introducing fresh variables
for them. 

State $\GraphInitCmpOne$ results from $\GraphInitCmpZero$ by evaluating the \code{load}
instruction at
$(\cmpFor{}, 0)$,
where the value \code{0}\linebreak stored at
$v_{\code{k\_ad}}$ is loaded to the program variable $\code{k}$. 
The next instruction is an \underline{i}nteger
\underline{c}o\underline{mp}ari\-son which checks whether the
\underline{u}nsigned value of \code{k} is \underline{l}ess \underline{t}han the one
of \code{n}. If we cannot\linebreak decide a comparison, we refine
the state into two successor states
like $\GraphInitCmpOneRefTrue$ and $\GraphInitCmpOneRefFalse$ 
(with
$(v_\code{n} > 0) \in\linebreak \KB^\GraphInitCmpOneRefTrue$ and
 $(v_\code{n} \leq 0) \in\KB^\GraphInitCmpOneRefFalse$). Evaluating
$\GraphInitCmpOneRefTrue$ yields $\code{kltn} = 1$ in $\GraphInitCmpTwoTrue$. Therefore, the 
\underline{br}anch instruc\-tion leads to the block \bodyFor{} in State
$\GraphInitBodyZero$. State $\GraphInitCmpOneRefFalse$ is evaluated to a state with
$\code{kltn} = 0$. This path branches to the block \initPtr{} and terminates as
\code{tail\_ptr} points to an empty list.

The instruction at $(\bodyFor{}, 0)$ allocates 16 bytes of memory
starting at $v_{\code{mem}}$
in State $\GraphInitBodyOne$.
Next, the pointer to the allocation is cast from \code{i8*} to
\code{list*} and assigned to \code{curr}. 
Now the\linebreak allocated area can be treated as a list element.
Then \code{nondet\_uint()} assigns a 32-bit integer to \code{nondet}.
The \code{getelementptr} instruction computes the address of the integer field of the list
element by indexing this field (the second \code{i32 0}) based on the start address
(\code{curr}). Since 
the address of this integer value coincides with the start address of
the list element, this instruction assigns $v_\code{mem}$ to \code{curr\_val}. 
Afterwards, the value of \code{nondet} is stored at \code{curr\_val} ($v_{\code{mem}}
\pointsto[\code{i32}] v_\code{nd}$), the value \code{0} stored at $v_\code{tp}$ is loaded
to \code{tail}, and a second \code{getelementptr} instruction computes the address of the
recursive field of the current list element ($v_{\code{cn}} = v_{\code{mem}} + 8$) and
assigns it to \code{curr\_next}, leading to 
State $\GraphInitBodySix$.
In the path to $\GraphInitBodyTen$,  the values of \code{tail} and \code{curr} are stored
at \code{curr\_next} and \code{tail\_ptr},  respectively ($v_{\code{cn}}
\pointsto[\code{list*}] 0$, $v_{\code{tp}} \pointsto[\code{list*}]
v_{\code{mem}}$). Finally, the incremented value of \code{k} is assigned to \code{kinc}
and stored at \code{k\_ad} ($v_{\code{k\_ad}} \pointsto[\code{i32}] 1$).

To ensure a finite graph construction, when a program position is reached for the second
time, we try to merge the states at this position to a \emph{generalized}
state. However, this is only meaningful if the domains of the $\LV$ functions
 of the two states coincide (i.e., the states consider the same program
variables). Therefore, after the branch from the loop body back to  \cmpFor{}
(State $\GraphInitCmpZeroA$ in Fig.\ \ref{fig:ForLoopSecondIteration}), we evaluate
the loop a
second time and reach $\GraphInitCmpZeroB$. Here, a second
\def\distB{0.85cm}
\begin{wrapfigure}[4]{r}{4.2cm}
\vspace*{-0.5cm}
\begin{tikzpicture}[node distance=2cm, auto]
\hspace*{-0.15cm}
\node (CmpZeroA) at (-0.67,0.07) {\small{$\GraphInitCmpZeroA:$}};
\node[label={[label distance=-0.55cm]0:{\scriptsize{$v_\codevar{mem}$}}}]  (vmem) at (0,0) {};
\node[data, right=0.35cm of vmem] (L1A) {\scriptsize{$v_\codevar{nd}$} \nodepart{second} \scriptsize{0}};
   
\node[above=0.17cm of $(L1A.west)!0.5!(L1A.text split)$] (o1L1A) {\tiny{\code{value}}};
\node[above=0.17cm of $(L1A.east)!0.5!(L1A.text split)$] (o2L1A) {\tiny{\code{next}}};

\path[ptr]  ($(vmem)+(0.25,0)$) --++(right:2mm)  |- (L1A.text west);

\node (CmpZeroB) at ($(-0.67,0.07)+(0,-\distB)$) {\small{$\GraphInitCmpZeroB:$}};
\node[] (distanceB) at (0,-\distB) {};
\node[label={[label distance=-0.55cm]0:{\scriptsize{$w_\codevar{mem}$}}}]  (wmem) at (distanceB) {};
\node[data, right=0.35cm of wmem] (L1B) {\scriptsize{$w_\codevar{nd}$} \nodepart{second}};
\node[data, right=0.2cm of L1B]   (L2B) {\scriptsize{$v_\codevar{nd}$} \nodepart{second} \scriptsize{0}};
   
\node[above=0.17cm of $(L1B.west)!0.5!(L1B.text split)$] (o1L1B) {\tiny{\code{value}}};
\node[above=0.17cm of $(L1B.east)!0.5!(L1B.text split)$] (o2L1B) {\tiny{\code{next}}};
\node[above=0.17cm of $(L2B.west)!0.5!(L2B.text split)$] (o1L2B) {\tiny{\code{value}}};
\node[above=0.17cm of $(L2B.east)!0.5!(L2B.text split)$] (o2L2B) {\tiny{\code{next}}};

\path[ptr]  ($(wmem)+(0.25,0)$) --++(right:2mm)  |- (L1B.text west);
\draw[fill] ($(L1B.east)!0.5!(L1B.text split)$) circle (0.05);
\draw[ptr]  ($(L1B.east)!0.5!(L1B.text split)$) --++(right:2mm) |- (L2B.text west);

\end{tikzpicture}
\vspace*{-.1cm}
\end{wrapfigure}
 list element with 
value
$w_\code{nd}$ and a \code{next} pointer $w_\code{cn}$ pointing to $v_\code{mem}$ has been
stored at a new allocation
$\llbracket{}w_\code{mem}, w_\code{mem}^\mathit{end}\rrbracket$. Now, \code{curr} points to
the new element and \code{k} has been incremented again, so \code{k\_ad} points to
2.
  
\input{graph_init_second}

We want to merge $\GraphInitCmpZeroA$ and
  $\GraphInitCmpZeroB$ to a more general state $\GraphInitCmpZeroGen$ that repre\-sents all states which are
represented by  $\GraphInitCmpZeroA$ or
  $\GraphInitCmpZeroB$.
The challenging part during generalization is to find loop invariants
 that provide sufficient information to
  prove termination of the loop.
For $\GraphInitCmpZeroGen$, we can neither use 
that \code{curr} points to a struct whose \code{next} field contains the null pointer (as
  in $\GraphInitCmpZeroA$), nor that its \code{next} field points to another struct
  whose \code{next} field contains the null pointer (as in $\GraphInitCmpZeroB$).
  
We solve this problem by introducing \emph{list invariants}. In our
example, we will infer an invariant stating that \code{curr} points to a list of
length $x_\ell \geq 1$. This invariant also implies that all struct fields are allocated
  and that there is no sharing.

To merge two states $s$ and $s'$ at the same program position with $\domain(\LV^s) =
\domain(\LV^{s'})$, we introduce a fresh symbolic variable $x_\codevar{var}$ for each program
variable \codevar{var} and use instantiations $\mu_\StateA$ and $\mu_\StateB$ which map
$x_\codevar{var}$ to the corresponding symbolic variables of $\StateA$ and
$\StateB$. For the merged state $\StateGen$, we set
$\LV^\StateGen(\codevar{var}) = x_\codevar{var}$. Moreover, we identify corresponding
variables that
only occur in the memory components and extend  $\mu_\StateA$ and $\mu_\StateB$ accordingly.
In a second step, we check which constraints from the memory components and the knowledge
base hold in both states in order to find invariants that we can add to the memory
components and the knowledge base of $\StateGen$. For example, if
$\alloc{\mu_\StateA(x)}{\mu_\StateA(x^{end})} \in \AL^\StateA$ and
$\alloc{\mu_\StateB(x)}{\mu_\StateB(x^{end})} \in \AL^\StateB$ for $x, x^{end} \in \Vsym$,
then $\alloc{x}{x^{end}}$ is added to $\AL^\StateGen$.  To extend this heuristic
to lists, we have to regard several memory entries together. 
If there is an
$\codevar{ad} \in \Ids$ such that $\mu_\StateA(x_\codevar{ad})$ and
$\mu_\StateB(x_\codevar{ad})$ both point to lists of type \codevar{ty}
but of different lengths $\ell_\StateA \neq
\ell_\StateB$ with $\ell_\StateA, \ell_\StateB \geq 1$, then we create a list invariant.

In our example, $\GraphInitCmpZeroA$
and $\GraphInitCmpZeroB$ 
contain lists of length $\ell_\GraphInitCmpZeroA =
1$ and $\ell_\GraphInitCmpZeroB = 2$.
Thus, when merging $\GraphInitCmpZeroA$ and $\GraphInitCmpZeroB$ to a new state
$\GraphInitCmpZeroGen$ (see \cref{fig:ForLoopMerging}), the lists are merged to a list invariant of variable length
$x_\ell$ and our technique
adds the formulas $1 \leq x_\ell$ and $x_\ell = x_\code{kinc}$
to $\KB^\GraphInitCmpZeroGen$. 
Moreover, the \codevar{i32} value of the first element is identified with
$x_\code{nd}$, since $\mu_\GraphInitCmpZeroA(x_\code{nd})$ is equal to the first value of
the first list element in $\GraphInitCmpZeroA$ and $\mu_\GraphInitCmpZeroB(x_\code{nd})$ is
equal to the first value of the first list element in $\GraphInitCmpZeroB$. Similarly, the
values of the last list elements are identified
  with $0$, as in $\GraphInitCmpZeroA$ and $\GraphInitCmpZeroB$.

After merging $s$ and $s'$, we
continue symbolic execution from the generalized state $\overline{s}$.
 The next time we reach the same program position, we might have to merge
the corresponding states again. As described in \cite{LLVM-JAR17}, we use
a heuristic  which ensures
that after a finite number of iterations, a state is reached
that is represented by an \emph{already existing} state in the SEG. Then symbolic execution can continue from this more general state
instead. So 
the construction always ends in a complete SEG or  an SEG containing the state $\ERROR$.

To prove termination of a program $\Prog$,  as
 in \cite{LLVM-JAR17} the cycles of the SEG are
translated to an ITS whose termination implies termination of $\Prog$.
 In our example, the first cycle of the SEG (corresponding to the \code{for} loop of
the \tool{C} program) yields transitions which terminate
since \code{k} is increased until it reaches
\code{n}.
The second cycle (corresponding to the \code{while} loop)
terminates since the length of the list decreases. Although there is no program
variable for the length, due to our list invariants the states of the SEG
contain variables for this length, which are also passed
to the ITS.  Thus, the  resulting transitions  clearly terminate.

\input{graph_init_merge}

%% file: graph_init_first.tex
\footnotesize
\newcommand{\ydist}{0.3cm}
\newcommand{\xdist}{0.5cm}
\newcommand{\setfont}{\scriptsize}
\newcommand{\labelxshift}{-1pt}

\begin{figure}[t]
\vspace*{-.2cm}
\centering
\begin{tikzpicture}[node distance = \ydist and \xdist]
\scriptsize
\def\widetwidth{6.2cm}
\def\smalltwidth{4.75cm}
\def\fulllinewidhth{11cm}
\def\edgenodedist{0.09cm}
\def\FirstIndentwidth{3cm}
\def\SecondIndentwidth{2cm}
\tikzset{invisible/.style={opacity=0}}
\tikzstyle{state}=[
                           inner sep=2pt,
                           font=\scriptsize,
                           draw]

\node[state, label={[xshift=\labelxshift]2:$\GraphInitEntryZero$}] (1) 
{$(\code{entry}, 0),\;
  \emptyset, \;
  \emptyset, \;
  \emptyset, \;
  \emptyset, \;
  \emptyset
 $
};

\node[state, below=of 1, align=left, label={[xshift=\labelxshift]2:$\GraphInitCmpZero$}] (2) 
{$(\cmpFor, 0),\;
  \{ \code{n} = v_{\code{n}}, \,
     \code{tail\_ptr} = v_{\code{tp}}, \,
     \code{k\_ad} = v_{\code{k\_ad}}, \,
     ...\}, \;
  \{ \alloc{v_{\code{tp}}}{v_{\code{tp}}^\mathit{end}}, \,
     \alloc{v_{\code{k\_ad}}}{v_{\code{k\_ad}}^\mathit{end}} \},
 $\\
 $
  \{ v_{\code{tp}} \pointsto[\code{list*}] 0, \,
     v_{\code{k\_ad}} \pointsto[\code{i32}] 0 \}, \;
  \emptyset, \;
  \{ v_{\code{tp}}^\mathit{end} = v_{\code{tp}} + 7, \,
     v_{\code{k\_ad}}^\mathit{end} = v_{\code{k\_ad}} + 3, \,
     ... \}
 $
};

\node[state, below=of 2, align=left, label={[xshift=\labelxshift]2:$\GraphInitCmpOne$}] (3) 
{$(\cmpFor, 1),\;
  \{ \code{k} = 0, \,
     ...\}, \;
  \AL^{\GraphInitCmpZero}, \;
  \PT^{\GraphInitCmpZero}, \;
  \emptyset, \;
  \KB^{\GraphInitCmpZero}
 $
};

\node[state, below left=.3cm and -2cm of 3, align=left, label={[xshift=\labelxshift]2:$\GraphInitCmpOneRefTrue$}] (3a) 
{$(\cmpFor, 1),\;
  \{ \code{k} = 0, \,
     ...\}, \;
  \AL^{\GraphInitCmpZero},
 $\\
 $
  \PT^{\GraphInitCmpZero}, \;
  \emptyset, \;
  \{ v_\code{n} > 0, \,
     ... \}
 $
};

\node[state, below right=.3cm and -2cm of 3, align=left, label={[xshift=\labelxshift]2:$\GraphInitCmpOneRefFalse$}] (3b) 
{$(\cmpFor, 1),\;
  \{ \code{k} = 0, \,
     ...\}, \;
  \AL^{\GraphInitCmpZero},
 $\\
 $
  \PT^{\GraphInitCmpZero}, \;
  \emptyset, \;
  \{ v_\code{n} \leq 0, \,
     ... \}
 $
};

\node[state, below right=.3cm and -4cm of 3a, align=left, label={[xshift=\labelxshift]2:$\GraphInitCmpTwoTrue$}] (4a) 
{$(\cmpFor, 2),\;
  \{ \code{kltn} = 1, \,
     ...\}, \;
  \AL^{\GraphInitCmpZero}, \;
  \PT^{\GraphInitCmpZero}, \;
  \emptyset, \;
  \KB^{\GraphInitCmpOneRefTrue}
 $
};

\node[below=of 3b] (4b) 
{$...$};

\node[state, below right=.3cm and -4.5cm of 4a, align=left, label={[xshift=\labelxshift]2:$\GraphInitBodyZero$}] (5) 
{$(\bodyFor, 0),\;
  \LV^{\GraphInitCmpTwoTrue}, \;
  \AL^{\GraphInitCmpZero}, \;
  \PT^{\GraphInitCmpZero}, \;
  \emptyset, \;
  \KB^{\GraphInitCmpOneRefTrue}
 $
};

\node[state, below right=.3cm and -6.5cm of 5, align=left, label={[xshift=\labelxshift]2:$\GraphInitBodyOne$}] (6) 
{$(\bodyFor, 1),\;
  \{ \code{mem} = v_{\code{mem}}, \,
     ...\}, \;
  \{ \alloc{v_{\code{mem}}}{v_{\code{mem}}^\mathit{end}}, \,
     ... \}, \;
  \PT^{\GraphInitCmpZero}, \;
  \emptyset, \;
  \{ v_{\code{mem}}^\mathit{end} = v_{\code{mem}} + 15, \,
     ... \}
 $
};

\node[state, below=of 6, align=left, label={[xshift=\labelxshift]2:$\GraphInitBodySix$}] (11) 
{$(\bodyFor, 7),\;
  \{ \code{curr} = v_{\code{mem}}, \,
     \code{nondet} = v_{\code{nd}}, \,
     \code{curr\_val} = v_{\code{mem}}, \,
     \code{tail} = 0, \,
     \code{curr\_next} = v_{\code{cn}}, \,
     ...\},$\\
 $\AL^{\GraphInitBodyOne}, \;
  \{ v_{\code{mem}} \pointsto[\code{i32}] v_{\code{nd}}, \,
     ... \}, \;
  \emptyset, \;
  \{ v_{\code{cn}} = v_{\code{mem}} + 8, \,
     ... \}
 $
};

\node[state, below=of 11, align=left, label={[xshift=\labelxshift]2:$\GraphInitBodyTen$}] (15) 
{$(\bodyFor, 11),\;
  \{ \code{kinc} = 1, \;
     ...\}, \,
  \AL^{\GraphInitBodyOne}, \;
  \{ v_{\code{cn}} \pointsto[\code{list*}] 0, \,
     v_{\code{tp}} \pointsto[\code{list*}] v_{\code{mem}}, \,
     v_{\code{k\_ad}} \pointsto[\code{i32}] 1, \,
     ... \}, \;
  \emptyset, \;
  \KB^{\GraphInitBodySix}
 $
};

\draw[omit-edge] (1)  --  (2);
\draw[eval-edge] (2)  --  (3);
\draw[eval-edge] (3)  --  (3a);
\draw[eval-edge] (3)  --  (3b);
\draw[eval-edge] (3a) --  (4a);
\draw[eval-edge] (3b) --  (4b);
\draw[eval-edge] (4a) --  (5);
\draw[eval-edge] (5)  --  (6);
\draw[omit-edge] (6)  --  (11);
\draw[omit-edge] (11) --  (15);

\end{tikzpicture}
\caption{SEG for the First Iteration of the \code{for} Loop}
\label{fig:ForLoopFirstIteration}
\vspace*{-.2cm}
\end{figure}
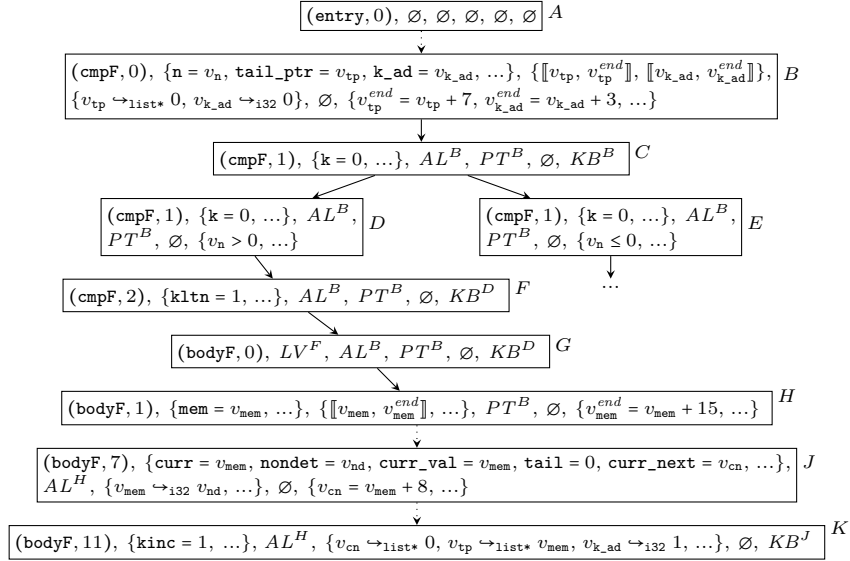
\normalsize

%% file: graph_init_second.tex
\footnotesize

\begin{figure}[t]
\vspace*{-.2cm}
\centering
\begin{tikzpicture}[node distance = \ydist and \xdist]
\scriptsize
\def\widetwidth{6.2cm}
\def\smalltwidth{4.75cm}
\def\fulllinewidhth{11cm}
\def\edgenodedist{0.2cm}
\def\FirstIndentwidth{3cm}
\def\SecondIndentwidth{2cm}
\tikzset{invisible/.style={opacity=0}}
\tikzstyle{state}=[
                           inner sep=2pt,
                           font=\scriptsize,
                           draw]

\node[state, align=left, label={[xshift=-\labelxshift]2:$\GraphInitCmpZeroA$}] (1) 
{$(\cmpFor, 0),\;
  \{
   \code{n} = v_{\code{n}}, \,
   \code{tail\_ptr} = v_{\code{tp}}, \,
   \code{mem} = v_{\code{mem}}, \,
   \code{curr} = v_{\code{mem}}, \,
   \code{nondet} = v_{\code{nd}}, \,
   \code{curr\_val} = v_{\code{mem}},$\\ 
   $\,\code{curr\_next} = v_{\code{cn}}, \,
\code{k} = 0, \,
   \code{kinc} = 1, \,
   ...\},\;
  \{
   \alloc{v_{\code{tp}}}{v_{\code{tp}}^\mathit{end}}, \,
   \alloc{v_{\code{k\_ad}}}{v_{\code{k\_ad}}^\mathit{end}}, \,
   \alloc{v_{\code{mem}}}{v_{\code{mem}}^\mathit{end}}
   \},$\\
 $\{
   v_{\code{tp}} \pointsto[\code{list*}] v_{\code{mem}}, \,
   v_{\code{k\_ad}} \pointsto[\code{i32}] 1, \,
   v_{\code{mem}} \pointsto[\code{i32}] v_{\code{nd}}, \,
   v_{\code{cn}} \pointsto[\code{list*}] 0
   \}, \;
  \emptyset,$\\
 $\{ v_{\code{n}} > 0, \,
    v_{\code{k\_ad}}^\mathit{end} = v_{\code{k\_ad}} + 3, \,
   v_{\code{tp}}^\mathit{end} = v_{\code{tp}} + 7, \,
   v_{\code{mem}}^\mathit{end} = v_{\code{mem}} + 15, \,
   v_{\code{cn}} = v_{\code{mem}} + 8, \,
   ... \}$
};

\node[state, below=of 1, align=left, label={[xshift=-\labelxshift]2:$\GraphInitCmpZeroB$}] (2) 
{$(\cmpFor, 0),\;
  \{   
   \code{n} = v_{\code{n}}, \,
   \code{tail\_ptr} = v_{\code{tp}}, \,
   \code{mem} = w_{\code{mem}}, \,
   \code{curr} = w_{\code{mem}}, \,
   \code{nondet} = w_{\code{nd}}, \,
   \code{curr\_val} = w_{\code{mem}},$\\
   $\,\code{curr\_next} = w_{\code{cn}}, \,
\code{k} = 1, \,
   \code{kinc} = 2, \,
   ...\},\;
  \{
   \alloc{v_{\code{tp}}}{v_{\code{tp}}^\mathit{end}}, \,
   \alloc{v_{\code{k\_ad}}}{v_{\code{k\_ad}}^\mathit{end}}, \,
   \alloc{v_{\code{mem}}}{v_{\code{mem}}^\mathit{end}}, \,
   \alloc{w_{\code{mem}}}{w_{\code{mem}}^\mathit{end}}
   \},$\\
 $\{
   v_{\code{tp}} \pointsto[\code{list*}] w_{\code{mem}}, \,
   v_{\code{k\_ad}} \pointsto[\code{i32}] 2, \,
   v_{\code{mem}} \pointsto[\code{i32}] v_{\code{nd}}, \,
   v_{\code{cn}} \pointsto[\code{list*}] 0, \,
   w_{\code{mem}} \pointsto[\code{i32}] w_{\code{nd}}, \,
   w_{\code{cn}} \pointsto[\code{list*}] v_{\code{mem}}
   \}, \;
  \emptyset,$\\
 $\{v_{\code{n}}> 1, 
  v_{\code{k\_ad}}^\mathit{end} = v_{\code{k\_ad}}\!+\!3, 
   v_{\code{tp}}^\mathit{end} = v_{\code{tp}}\!+\!7,  
   v_{\code{mem}}^\mathit{end} = v_{\code{mem}}\!+\!15, 
   v_{\code{cn}} = v_{\code{mem}}\!+\!8, 
   w_{\code{mem}}^\mathit{end} = w_{\code{mem}}\!+\!15, 
   w_{\code{cn}} = w_{\code{mem}}\!+\!8, 
   ... \}$
};

\draw[omit-edge] (1) -- (2);

\end{tikzpicture}
\caption{Second Iteration of the \code{for} Loop}
\label{fig:ForLoopSecondIteration}
\vspace*{-.2cm}
\end{figure}
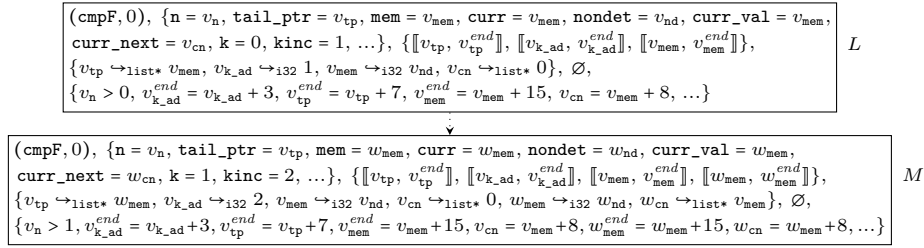
\normalsize

%% file: graph_init_merge.tex
\footnotesize

\begin{figure}[t]
\vspace*{-.3cm}
\centering
\begin{tikzpicture}[node distance = \ydist and \xdist]
\scriptsize
\def\widetwidth{6.2cm}
\def\smalltwidth{4.75cm}
\def\fulllinewidhth{11cm}
\def\edgenodedist{0.2cm}
\def\FirstIndentwidth{3cm}
\def\SecondIndentwidth{2cm}
\tikzset{invisible/.style={opacity=0}}
\tikzstyle{state}=[
                           inner sep=2pt,
                           font=\scriptsize,
                           draw]

\node[state, align=left, label={[xshift=-\labelxshift]91:$\GraphInitCmpZeroA$}] (1) {};

\node[state, below=of 1, align=left, label={[xshift=-\labelxshift]269:$\GraphInitCmpZeroB$}] (2) {};

\node[state, right=of 1, yshift=-.23cm, xshift=.5cm, align=left, label={[xshift=-\labelxshift]175:$\GraphInitCmpZeroGen$}] (3) 
{$(\cmpFor, 0),\;
  \{
     \code{n} = x_{\code{n}}, \,
     \code{tail\_ptr} = x_{\code{tp}}, \,
   \code{mem} = x_{\code{mem}}, \,
   \code{curr} = x_{\code{mem}}, \,
   \code{nondet} = x_{\code{nd}}, \,
   \code{curr\_val} = x_{\code{mem}},$\\
$\,
   \code{curr\_next} = x_{\code{cn}}, \, \code{k} = x_{\code{k}}, \,
   \code{kinc} = x_{\code{kinc}}, \,
    ...\},\;
  \{
   \alloc{x_{\code{tp}}}{x_{\code{tp}}^\mathit{end}}, \,
   \alloc{x_{\code{k\_ad}}}{x_{\code{k\_ad}}^\mathit{end}}
   \},$\\
 $\{
   x_{\code{tp}} \pointsto[\code{list*}] x_{\code{mem}}, \,
   x_{\code{k\_ad}} \pointsto[\code{i32}] x_{\code{kinc}}
  \}, \;
  \{
   x_{\code{mem}} \pointstorec[\code{list}]{x_{\ell}}
     [(0: \code{i32}: x_{\code{nd}}..\hat{x}_{\code{nd}}),
      (8: \code{list*}: x_{\code{next}}..0)]
   \},$\\
 $\{
      x_{\code{n}} > x_{\code{k}}, \,
x_{\code{k\_ad}}^\mathit{end} = x_{\code{k\_ad}} + 3, \,
   x_{\code{tp}}^\mathit{end} = x_{\code{tp}} + 7, \,
   x_{\code{cn}} = x_{\code{mem}} + 8, \,
   x_{\code{kinc}} = x_{\code{k}} + 1, \,
   1 \leq x_\ell, \,
   x_{\ell} = x_{\code{kinc}}, \,
   ... \}$
};

\draw[omit-edge] (1) -- (2);

\coordinate (merge-p1) at ($(1.east)+(\edgenodedist*1.6,0)$);
\coordinate (merge-p2) at (merge-p1 |- 2.east);
\draw[-,rounded corners=3pt] (1.east) -- (merge-p1) -- (merge-p2) -- (2.east);

\coordinate (gen-p1) at ($(1.south east)+(\edgenodedist*1.6,-\ydist/2)$);
\draw[gen-edge] (gen-p1) --  (3.west);

\end{tikzpicture}
\caption{Merging of States}
\label{fig:ForLoopMerging}
\vspace*{-.2cm}
\end{figure}
\normalsize

%% file: evaluation.tex
\section{Conclusion and Evaluation}
\label{sect:evaluation}

We presented a new approach for proving memory safety and termination of
\tool{C} programs on lists.  The main idea is
to
extend the  states in the SEG by  \emph{list
invariants}. We 
developed techniques to infer and  modify list invariants
automatically during the symbolic execution.

List invariants 
abstract from a concrete number of memory alloca\-tions to a list of allocations of variable
length while preserving knowledge about some of the contents and the list\linebreak shape.
They also contain information on
the memory arrangement of the list fields which is\linebreak needed to prove memory safety for  programs that
access fields
via
pointer arithmetic.
The symbolic variables for  the list length and the first and last values of list
elements are preserved when generating an ITS from the SEG.
 Thus, they
  can be used in its termination proof.

In \cite{RTA11,CAV12,RTA10} we
developed a technique for
termination analysis of \tool{Java},
based on a program transformation to \emph{integer term rewrite systems} 
instead of ITSs.  This approach does not require specific
list
invariants as
recursive data structures on the heap are abstracted to terms. However, these terms are
unsuitable for \sfC{}, since they cannot express 
memory allocations and the connection to 
their contents.

For every abstract state, we define a corresponding \emph{separation logic} formula to
define which concrete states are represented by the abstract state, see \cite{report}.  To
extend this formula to list invariants, we use  
a fragment similar to
\emph{quantitative} separation logic \cite{QSL}, extending conventional separation logic
by list predicates.
Based on separation logic with inductive predicates,
\cite{Predator} also uses a representation of lists  with offsets which is similar to our
list invariants, and \cite{Calcagno06} presents an abstract domain to represent lists in the presence of
pointer arithmetic. However, in contrast to our work, \cite{Calcagno06,Predator} are not concerned with
termination analysis.

Separation logic predicates for termination of list programs were also used in
\cite{Mutant},  but 
their list predicates only consider the list length and
the recursive field,  but no other fields or offsets.  
The tools \tool{Cyclist} \cite{Cyclist} and \tool{HipTNT+} \cite{HipTNT} are integra\-ted
in separation logic systems which also allow to define heap predicates.
However, they require annotations and
hints which parameters of the list predicates are needed as a
termination measure.
The tool \tool{2LS} \cite{2LS} also provides basic support for dynamic data
structures.
But all these approaches are not suitable if
ter\-mination depends on the contents or the shape of data structures combined with\linebreak pointer
arithmetic.
In \cite{David15}, programs can be annotated with 
arithmetic and
structural properties to reason about termination.  In contrast,
our approach does not need hints or annotations, but finds termination arguments fully
automatically.

We implemented our approach in \aprove{} \cite{LLVM-JAR17}.
Since existing tools can hardly prove termination
of \sfC{} programs with  lists,
the current benchmarks for termination analysis contain almost no list
programs. In 2017, a 
set
of 18 \sfC{} programs
 on 
lists was added to the \emph{Termination} category of the
\emph{Competition on Software Verification}
(\emph{SV-COMP}),\footnote{\url{https://sv-comp.sosy-lab.org/2022/}} where nine of them are
terminating.
Two of these nine programs do not need
list invariants, because they
just  create
a list without operating on it
afterwards. The remaining seven terminating programs create a list and then
traverse it, search for a value, or append lists and compute the length
afterwards.
Our new version of \aprove{} is the only termination prover
that succeeds for five of these seven programs, since here
termination depends on the shape or contents of a list after its
creation.

For the \emph{Termination Competition} 2022,\footnote{\url{https://termination-portal.org/wiki/Termination_Competition_2022}}
we submitted 18
terminating \sfC{} programs on
lists (different
from the ones at \emph{SV-COMP}), where \aprove{} succeeds on 17 of them.
Two of these programs just create a list. Three  traverse it
afterwards (by a loop or recursion),
and ten search for a value, where
 for nine, also the list contents 
are relevant for termination.
Three programs perform common
operations like inserting or deleting an element.

Without list invariants, in each collection
\aprove{}  only proves termination of two
examples that just create a list without traversing it afterwards,
and non-termination for one example in the \emph{SV-COMP} collection.
 \aprove{}  and \tool{UAutomizer} were the most powerful \sfC{} termination tools 
 in  \emph{SV-COMP} and the \emph{Termination Competition} 2022, with \tool{UAutomizer}  winning the former  
\begin{wrapfigure}[3]{r}{8.3cm}
 \footnotesize
 \renewcommand{\arraystretch}{.9}
\vspace*{-.37cm}
 \begin{tabular}{|c|c|c|c|}
    \cline{2-4}
    \multicolumn{1}{c|}{} &  {\scriptsize \textbf{SV-C T.}} 
                          &  {\scriptsize \textbf{SV-C Non-T.}} 
                          &  {\scriptsize \textbf{TermCmp T.}}\\
    \hline
    \aprove{}             & 7 (of 9) & 5 (of 9) & 17 (of 18)\\
    \hline
    \tool{UAutomizer}     & 2 (of 9) & 7 (of 9) &  1 (of 18)\\
    \hline
   \end{tabular}
 \renewcommand{\arraystretch}{1}
\end{wrapfigure}
and \aprove{} winning the
latter. To download \aprove{},  its web interface, and 
 details on our experiments, see \url{https://aprove-developers.github.io/recursive_structs}.

 \vspace*{-.2cm}

%% file: paper.bbl
\providecommand{\noopsort}[1]{}\providecommand{\apro}{}\providecommand{\ecli}{}

%% file: paper.bbl
\begin{thebibliography}{10}

\bibitem{Mutant}
J.\ Berdine, B.\ Cook, D.\ Distefano, and P.~W.\ O'Hearn.
\newblock Automatic termination proofs for programs with shape-shifting heaps.
\newblock In {\em Proc.\ CAV~'06}, LNCS 4144, pages 386--400, 2006.

\bibitem{QSL}
M.\ Bozga, R.\ Iosif, and S.\ Perarnau.
\newblock Quantitative separation logic and programs with lists.
\newblock {\em J.\ Aut.\ Reasoning}, 45(2):131--156, 2010.

\bibitem{RTA11}
M.\ Brockschmidt\noopsort{4}, C.\ Otto, and J.\ Giesl.
\newblock Modular termination proofs of recursive {{\tool{Java Bytecode}}}
  programs by term rewriting.
\newblock In {\em Proc.\ RTA~'11}, LIPIcs 10, pages 155--170, 2011.

\bibitem{CAV12}
M.\ Brockschmidt\noopsort{5}, R.\ Musiol, C.\ Otto, and J.\ Giesl.
\newblock Automated termination proofs for {{\tool{Java}}} programs with cyclic
  data.
\newblock In {\em Proc.\ CAV~'12}, LNCS 7358, pages 105--122, 2012.

\bibitem{Calcagno06}
C.\ Calcagno, D.\ Distefano, P.~W.' O'Hearn, and H.\ Yang.
\newblock Beyond reachability: Shape abstraction in the presence of pointer
  arithmetic.
\newblock In {\em Proc.\ SAS~'06}, LNCS 4134, pages 182--203, 2006.

\bibitem{David15}
C.\ David, D.\ Kroening, M.\ Lewis, and J.\ Vitek.
\newblock Propositional reasoning about safety and termination of
  heap-manipulating programs.
\newblock In {\em Proc.\ ESOP~'15}, LNCS 9032, pages 661--684, 2015.

\bibitem{Predator}
K.\ Dudka, P.\ Peringer, and T.\ Vojnar.
\newblock {{\tool{{Predator}}}}: {A} practical tool for checking manipulation
  of dynamic data structures using separation logic.
\newblock In {\em Proc.\ CAV~'11}, LNCS 6806, pages 372--378, 2011.

\bibitem{ASV19}
F.\ Emrich, J.\ Hensel, and J.\ Giesl.
\newblock {{\textsf{AProVE}}}: Modular termination analysis of
  memory-manipulating {{\textsf{C}}} programs.
\newblock {\em CoRR}, abs/2302.02382, 2023.

\bibitem{JAR17-AProVE}
J.\ Giesl, C.\ Aschermann, M.\ Brockschmidt, F.\ Emmes, F.\ Frohn, C.\ Fuhs,
  C.\ Otto, M.~Pl\"ucker, P.\ Schneider-Kamp, T.\ Str\"oder, S.\ Swiderski, and
  R.\ Thiemann.
\newblock Analyzing program termination and complexity automatically with
  \textsf{AProVE}.
\newblock {\em J.\ Aut.\ Reasoning}, 58(1):3--31, 2017.

\bibitem{JLAMP18}
J.\ Hensel, J.\ Giesl, F.\ Frohn, and T.\ Str\"oder.
\newblock Termination and complexity analysis for programs with bitvector
  arithmetic by symbolic execution.
\newblock {\em J.\ Log.\ Alg.\ Meth.\ Prog.}, 97:105--130, 2018.

\bibitem{report}
J.\ Hensel\noopsort{1} and J.\ Giesl.
\newblock Proving termination of {{\textsf{C}}} programs with lists.
\newblock In {\em Proc.\ CADE~'23}, LNCS, 2023.
\newblock To appear. Full version appeared in \emph{CoRR}, abs/2305.12159.

\bibitem{HipTNT}
T.~C.\ Le, S.\ Qin, and W.\ Chin.
\newblock Termination and non-termination specification inference.
\newblock In {\em Proc.\ PLDI~'15}, pages 489--498, 2015.

\bibitem{2LS}
V.\ Mal{\'i}k, {\v{S}}.\ Marti{\v{c}}ek, P.\ Schrammel, M.\ Srivas, T.\ Vojnar,
  and J.\ Wahlang.
\newblock \textsf{2LS}: Memory safety and non-termination.
\newblock In {\em Proc.\ TACAS~'18}, LNCS 10806, pages 417--421, 2018.

\bibitem{RTA10}
C.\ Otto, M.\ Brockschmidt, C.\ von Essen, and J.\ Giesl.
\newblock Automated termination analysis of \tool{Java Bytecode} by term
  rewriting.
\newblock In {\em Proc.\ RTA~'10}, LIPIcs 6, pages 259--276, 2010.

\bibitem{Cyclist}
R.~N.~S.\ Rowe and J.\ Brotherston.
\newblock Automatic cyclic termination proofs for recursive procedures in
  separation logic.
\newblock In {\em Proc.\ CPP~'17}, pages 53--65, 2017.

\bibitem{LLVM-JAR17}
T.\ Str\"oder, J.\ Giesl, M.\ Brockschmidt, F.\ Frohn, C.\ Fuhs, J.\ Hensel,
  P.\ Schneider-Kamp, and C.\ Aschermann.
\newblock Automatically proving termination and memory safety for programs with
  pointer arithmetic.
\newblock {\em J.\ Aut.\ Reasoning}, 58(1):33--65, 2017.

\end{thebibliography}
